MPI–PhT/95-119

# Improvements of Lüscher's local bosonic fermion algorithm

Beat Jegerlehner[a]

[a] Max-Planck-Institut für Physik
Föhringer Ring 6
D-80805 München, Germany

**Abstract**

We discuss the application of hybrid over-relaxation, even-odd preconditioning and a modified local updating procedure to the local bosonic fermion algorithm. Studies on autocorrelation times and the tuning of the parameters of the algorithm are done on various lattice sizes simulating $SU(2)$-LGT with 2 flavours of Wilson quarks. A substantial decrease of the computational cost could be achieved.

December 1995

# 1 Introduction

The incorporation of fermionic degrees of freedom in lattice simulations is a very general and longstanding algorithmical problem. Since there is no efficient stochastic way for calculating the determinant of the fermion matrix, one usually calculates the inverse determinant of the inverse fermion matrix, which introduces a non–local action. This limits the choice of algorithms in an essential way and so the computational effort for simulating fermions is about a factor 100–1000 larger than for bosons. The algorithm most commonly used today is the hybrid monte carlo algorithm (HMC) [13], which has a much worse volume $V$ and correlation length $\xi$ behaviour than local algorithms for bosonic fields.

In particular in lattice QCD this has made simulations including the effect of light sea-quarks extremely difficult. Therefore simulations are usually done with setting the fermion determinant to 1 (quenching), thus taking only into account the effects of the valence quarks. While this may be a fair approximation for some observables such as some aspects of the mass spectrum, considerable effects are expected for example for $\alpha_S$.

It seems therefore crucial to find new simulation methods or to improve on the existing fermion algorithms. One algorithm which is in the course of development and testing is the local bosonic algorithm introduced by M. Lüscher [1]. This paper will describe some improvements of this algorithm as well as a first study of its behaviour on lattice sizes up to $16^4$.

# 2 The local bosonic fermion algorithm

Since the algorithm was already discussed in great detail in [1,3], we will give only a short introduction and basic notations.

For the $SU(2)$-LGT with two flavours of Wilson quarks, the partition function is given by

$$Z_{QCD} = \int d[U] e^{-S_W[U]} \det(M^\dagger M), \tag{2.1}$$

with $M = 1 - KH$ the Wilson fermion matrix and $S_W[U]$ the usual Wilson gauge action. All simulations were done with periodic boundary conditions. For convenience, we work with $Q = c_0 \gamma_5 M$ with $c_0^{-1} = c_M(1+8K)$ instead of $M$. For $c_M \geq 1$ we have $||Q|| \leq 1$ and furthermore $Q^\dagger = Q$. The method is based on a polynomial approximation of the inverse of the fermion matrix. We employ here, as in [3], Chebychev polynomials to construct polynomials $P_n(x)$ of even degree $n$ which statisfy

$$|x P_n(x) - 1| \leq \delta \quad \forall x \in [\epsilon, 1], \tag{2.2}$$

where $\delta$ is given by

$$\delta = 2 \left( \frac{1-\sqrt{\epsilon}}{1+\sqrt{\epsilon}} \right)^{n+1}. \tag{2.3}$$

This means that, keeping $\epsilon$ fixed, we have to scale $n \propto -\ln(\delta)$ and, keeping $\delta$ fixed, $n \propto 1/\sqrt{\epsilon}$. Different possible polynomials are discussed in [1,9,12]. One then writes

$$\det Q^2 \approx (\det P_n(Q^2))^{-1} \propto \prod_{k=1}^{n} (Q - r_k)(Q - r_k^*) \qquad (2.4)$$

where $r_k$ is the square root of the $k$-th root of $P_n$ with Im $r_k > 0$. Introducing $n$ bosonic auxiliary fields, we can rewrite

$$Z_{QCD} \approx Z_B = \int d[U] e^{-S_W[U]} \prod_{k=1}^{n} d[\phi_k^\dagger] d[\phi_k] e^{-|(Q-r_k)\phi_k|^2}. \qquad (2.5)$$

We can now apply the standard local heatbath and over-relaxation procedures for both the bosonic fields and the gauge field. The first simulations have shown that the autocorrelation times are rather large [3,4] due to the fact that $n$ fields are coupled to one gauge field and are updated one after the other. This results in the fact that the bosonic fields are guided by the gauge field and vice versa, since they are not updated simultaneously. This guidance results in small step sizes and therefore the autocorrelation time is found to be proportional to $n$. On the other hand it was found that surprisingly large $\delta$, of the order of a few percent on lattices up to $6^3 \times 12$, still give good results. Also the tuning of $\epsilon$ seemed not to be critical.

## 3 Hybrid Over-Relaxation

Hybrid over-relaxation is the state of the art algorithm for bosonic systems when no cluster algorithm is available [8]. It is believed that this algorithm, which consists simply in the mixing of heatbath and over-relaxation sweeps with a ratio $1 : n_o$, has a dynamical critical exponent $z \approx 1$ if $n_o$ is tuned proportionally to $\xi$. This has been verified for the case of $SU(3)$ pure gauge theory.

In [4] it was discussed that application of hybrid over-relaxation to bosonic and gauge fields separately fails, since for fixed bosonic fields the gauge fields have a very narrow distribution around the bosonic force and vice versa. However, applying 1 bosonic heatbath sweep and $n_o$ pairs of gauge and bosonic over-relaxation sweeps a substantial improvement can be observed. (The gauge field needs not to be updated with a heatbath sweep since ergodicity is ensured by the bosonic heatbath update.)

In table 1 the autocorrelation time dependence is displayed for several updating schemes. Note that these numbers are obtained using the preconditioned algorithm, which will be discussed in the next section. The autocorrelation times are given in units of matrix-vector multiplications of the fermion matrix $Q$ (see also appendix C). It is seen that the autocorrelation time in units of $Q\phi$-multiplications can, in comparison to the original version of the update used in [3], be reduced from $\approx 45 \times 10^3$ by a factor of 3 to $\approx 15 \times 10^3$ in the 30-field case. When adding more over-relaxation sweeps $\tau_{\rm int}$ starts to rise again for the examined parameters. However, we expect the gain to increase with increasing correlation length.

| Updating | $\langle\Box\rangle$ | $\tau_{\rm int}(\Box)$ | $\tau_{\rm int}(C_\pi(T/2))$ |
|---|---|---|---|
| HOh | 0.4707(15) | 45(10) | 26(4) |
| HOo | 0.4701(12) | 30(8) | 15(3) |
| HoOo | 0.4708(14) | 20(4) | 12(2) |
| HOoOo | 0.4704(11) | 15(4) | 13(3) |
| HoOoOo | 0.4702(8) | 19(5) | 17(2) |
| HOh,g | 0.4690(19) | 34(11) | 29(8) |
| HOh,b | 0.4737(27) | 52(20) | 24(7) |
| HOh,gb | 0.4701(23) | 33(11) | 32(10) |
| HOh,GB | 0.4716(35) | 58(30) | 25(10) |
| HOoOo,gb | 0.4700(8) | 16(2) | 13(2) |

**Table 1:** Autocorrelation times in units of $1000 Q\phi$-multiplications on the $4^3 \times 8$ lattice for $\beta = 1.75$ and $K = 0.165$. All runs were done with $n = 30$, $\epsilon = 0.0045$ (thus $\delta \approx 0.03$) and $c_M = 1.1$. The letters in the first column give the type and order of sweeps used per iteration, where H is a bosonic heatbath, O a bosonic over-relaxation and h and o the gauge updates. A g denotes black-white checkerboarding of the gauge field, b the same applied to the bosonic fields. G and B denote the extended checkerboarding.

Generally we performed a sweep through the lattice as a series of lexicographically ordered local updates, i.e. by performing four loops over the four local coordinates. We also examined the effect of updating with checkerboard updates. We subdivided the lattice into 2 sublattices using the normal black-white checkerboarding and 16 sublattices with an extended checkerboard which were updated one after the other. For this, we coloured each site and link according to the functions

$$C(x) = (-1)^{x_1+x_2+x_3+x_4} \qquad \text{and} \tag{3.1}$$

$$C_e(x) = \frac{1}{2}\sum_{\mu=1}^{4} 2^{\mu-1}\left(1-(-1)^{x_\mu}\right). \tag{3.2}$$

We then updated one colour after the other. In the case of the extended checkerboard sites and links of the same colour do not interact with each other, i.e. they can be updated simultaneously. Due to the $Q^2$ term in the action this is not the case for the usual checkerboarding. In table 1 the results are presented. No significant improvement can be seen, which is in accordance with intuition, since for the gauge field the amount of change is mainly given by the bosonic force, which is held constant during the gauge update.

A further interesting possibility, namely to update gauge fields and bosonic fields one after the other locally, was not yet tested, since the present implementation of the algorithm does not allow to update all $n$ bosonic fields site by site efficiently. In this case the effect of checkerboarding might be much stronger.

## 4 Even-odd preconditioning

The number of fields needed in the approximation depends on the spectral condition number $k(Q) = \lambda_{\max}(Q)/\lambda_{\min}(Q)$ of the fermion matrix Moreover, it enters the autocorrelation time linearly [4]. It is well known that there are matrices with a determinant proportional to that of the fermion matrix which have lower condition number, so called preconditioned matrices. For the bosonic theory, however, it is necessary that the resulting action stays local. A simple and useful possibility which is known to work well is even-odd preconditioning [5]. One uses the fact that the off-diagonal part of the fermion matrix connects only even to odd sites and vice versa. If we write the fermion field in the following way:

$$\psi = \begin{pmatrix} \psi_e \\ \psi_o \end{pmatrix} \tag{4.1}$$

where $\psi_o$ denotes the part of the field which lives on the odd sites, we can write the fermion matrix $M$ as

$$M = \begin{pmatrix} 1 & -KD_{eo} \\ -KD_{oe} & 1 \end{pmatrix}. \tag{4.2}$$

The following identity proves to be very useful:

$$\det \begin{pmatrix} A & B \\ C & D \end{pmatrix} = \det A \det \left( D - CA^{-1}B \right). \tag{4.3}$$

Applying it to the fermion matrix, we get

$$\det M = \det \left( 1 - K^2 D_{oe} D_{eo} \right). \tag{4.4}$$

Let us denote the matrix on the right hand side with $\hat{M}$. This matrix connects only odd with odd sites, so the associated fermion fields live only on half of the lattice sites.

The eigenvalues of $\hat{M}$ are connected to the ones of $M$ via

$$\hat{\lambda} = 2\lambda - \lambda^2, \tag{4.5}$$

and $||\hat{M}||$ is bounded by $1 + 64K^2$ in contrast to $1 + 8K$ for $||M||$. Note that while $||M_0|| = 1 + 8K$, where $M_0$ is the free fermion matrix, $||\hat{M}_0||$ is generally smaller than $1 + 64K^2$. For the bosonic algorithm we use

$$\hat{Q} = \tilde{c}_0 \gamma_5 (1 - K^2 D_{oe} D_{eo}) \tag{4.6}$$

with $\tilde{c}_0^{-1} = c_M(1 + 64K^2)$, which fulfills $\hat{Q}^\dagger = \hat{Q}$ and $||\hat{Q}|| \leq 1$ for $c_M \geq 1$.

Unfortunately, $\hat{Q}^2$ connects (next to)$^3$ nearest neighbors so that a local update step becomes very complex. Fortunately, this problem can be circumvented by applying (4.3) once again *after* we applied the polynomial to $\hat{Q}^2$ [7]. We start from

$$P(\hat{Q}^2) = \prod_k (\hat{Q}^2 - z_k) = \prod_k (\hat{Q} - r_k)(\hat{Q} - r_k^*). \tag{4.7}$$

Now we can apply (4.3) to each factor and get

$$\det\left(\hat{Q}-r_k\right) \propto \det\begin{pmatrix} \tilde{c}_0\gamma_5 & -\tilde{c}_0\gamma_5 K D_{eo} \\ -\tilde{c}_0\gamma_5 K D_{oe} & \tilde{c}_0\gamma_5 - r_k \end{pmatrix}. \tag{4.8}$$

Letting $\tilde{Q} = \tilde{c}_0/c_0 Q$, we obtain the preconditioned action

$$\tilde{S}_b = \sum_k \phi_k^\dagger(\tilde{Q} - P_o r_k^*)(\tilde{Q} - P_o r_k)\phi_k, \tag{4.9}$$

which is very similar to the original one. $P_o$ denotes the projector on the odd sites. Explicit formulae for the updates are given in appendix A. One might wonder whether the reverse use of (4.3) does not destroy the advantages of the preconditioning, but this last transformation only affects the behaviour of the bosonic modes, which did not seem to be critical [4]. This is verified by the test results.

Note that the same trick also applies to other cases of preconditioning, for example if we use the inverse free fermion matrix, i.e. $\hat{M} = M_0^{-1}M$. $\hat{M}$ cannot easily be made hermitian, so that we have to use a non-hermitian approximation as described in [9]. We then have

$$P(\hat{M}) = \prod_k (\hat{M} - z_k)(\hat{M} - z_k^*) \tag{4.10}$$

and use

$$\begin{aligned} \det P(\hat{M}) &\propto \prod_k \det\left((M - z_k M_0)(M - z_k^* M_0)\right) \\ &= \prod_k \det\left((M - z_k M_0)(M^\dagger - z_k^* M_0^\dagger)\right), \end{aligned} \tag{4.11}$$

to get again a *local* action. In the last line we used the fact that $\gamma_5 M \gamma_5 = M^\dagger$. Since this kind of preconditioning is only expected to work well in the case of fixed gauge simulations, we did not test it. Note that for calculating the eigenvalues and for a global metropolis step, one still has to use and implement $\hat{M}$.

### 4.1 Numerical results

To be able to exploit the effect of the decreased condition number $k$ in the polynomial inversion, one has to study the lowest and highest eigenvalues of $\hat{Q}^2$ in detail. In figs. 1 and 2 we display the distributions of $\lambda_{\min}(\hat{Q}^2)$ and $\lambda_{\max}(\hat{Q}^2)$ as well as $\lambda_{\min}(\hat{Q}^2)/\lambda_{\min}(Q^2)$ and $k(Q^2)/k(\hat{Q}^2)$ for various lattice sizes ($c_M$ was set to 1). The $4^3\times 8$ lattice was run at $\beta = 1.75$ and $K = 0.165$, the other lattices at $\beta = 2.12$ and $K = 0.15$. The distribution of $\lambda_{\min}(\hat{Q}^2)/\lambda_{\min}(Q^2)$ as well as $k(Q^2)/k(\hat{Q}^2)$ is very narrow, showing that this kind of preconditioning works very well for the cases we investigated. With growing lattice size, $\lambda_{\min}(\hat{Q}^2)/\lambda_{\min}(Q^2)$ approaches the expected value of 3.25 for $K = 0.15$. Surprisingly, the condition number decreases by a factor of $\approx 8$ instead of the expected factor of $\approx 4$.

| $n, \epsilon$ | $c_M$ | Updating | $\langle\Box\rangle$ | $\langle\hat{\lambda}_{\min}\rangle$ | $\langle\hat{\lambda}_{\max}\rangle$ | $\tau_{\rm int}(\Box)$ |
|---|---|---|---|---|---|---|
| 0.0017 | 1.1 | HOh,non-pre | 0.4715(12) | 0.00321(8) | 0.6485(1) | 225(62) |
| 50 | | HOh | 0.4743(24) | 0.00614(18) | 0.2617(1) | 196(52) |
| 0.0045 | | HOh | 0.4707(15) | 0.00624(16) | | 45(10) |
| 30 | | HOoOo | 0.4704(11) | 0.00609(8) | | 15(4) |
| 0.015 | 0.6 | HOoOo | 0.4710(16) | 0.00612(11) | 0.8797(3) | 16(5) |
| 16 | 0.58 | HOoOo | 0.4692(13) | 0.00617(6) | 0.9414(3) | 23(5) |

**Table 2:** Autocorrelation times in units of $1000Q\phi$-multiplications on the $4^3 \times 8$ lattice for $\beta = 1.75$ and $K = 0.165$. $n$ and $\epsilon$ are chosen so that $\delta = 0.03$. All except the first run are preconditioned.

The reason is that $\lambda_{\max}(\hat{Q}^2)$ is far below the bound, growing slightly for larger lattice sizes. This was also seen for $SU(3)$ gauge configurations in [9]. This might indicate that the bound is actually too high. The widths of the distributions are roughly proportional to $1/\sqrt{V}$, as expected for a global quantity, and already very narrow for small lattice sizes, which simplifies the tuning of $\epsilon$ and $c_M$.

In [3] it was seen that observables were quite insensitive to $\epsilon$. We also expect this to be the case for the preconditioned case, since the argument was that $P(x)$ will still approximate $1/x$ even for $x < \epsilon$. In fig. 6 we investigated the dependence of several observables on $\epsilon$, keeping $\delta = 3\%$. We find, in agreement to our earlier results, that $\epsilon$ has little effect on the observables, as expected. The autocorrelation times seem to be consistent with a $\tau_{\rm int} \propto n/\sqrt{\epsilon}$ behaviour, as mentioned in [6]. One should note that the difference in the computational effort between the highest and lowest $\epsilon$ in fig. 6 is roughly a factor of 10.

In fig. 5 the dependence on $c_M$ was investigated, keeping $\epsilon$ and $\delta$ fixed. The observables are very sensitive to $\lambda_{\max}(\hat{Q}^2)$ when it gets larger than one. The plaquette is, as expected, most sensitive, but also the masses show significant deviations. The autocorrelations grow rapidly when $\lambda_{\max}(\hat{Q}^2)$ gets close to one. The reason is that there are always roots $r_k$ close to 1, so that $|\hat{Q} - r_k|^2$ develops small eigenvalues. Thus one expects that the corresponding bosonic field with the action $|(\tilde{Q} - P_o r_k)\phi_k|^2$ also develops slow modes. It is therefore necessary to monitor $\lambda_{\max}(\hat{Q}^2)$ and make sure that it is reasonably below 1.

We now come to the autocorrelation times of the preconditioned algorithm. In table 2, various runs for different $c_M$ with fixed $\delta$ are shown. The number of independent configurations used was around 200 for each run. The first line contains the original, non-preconditioned run. Using the same polynomial in the preconditioned case shows no significant effect on $\tau_{\rm int}$. We conclude that the bosonic modes are well behaved and the back-transformation (4.8) does not spoil the dynamics of the fields. Adjusting $\epsilon$ to the new minimal eigenvalue, keeping $\epsilon/\lambda_{\min}$ constant decreases the computational effort by a factor of $\approx 4.5$. By setting $c_M = 0.6$ and adjusting the polynomial accordingly, the computational effort does not decrease any more, although $n$ was decreased by a factor of 2. It looks like $\lambda_{\max}(\hat{Q}^2)$ is already too large at this $c_M$. A more detailed analysis is shown in fig. 4, where, keeping $\delta$ and $\epsilon/\lambda_{\min}(\hat{Q}^2)$ constant, different $c_M$ values were simulated. There seems to be an optimal value for $c_M$, coming from a trade-off between $n$ and the slow mode coming from $\lambda_{\max}(\hat{Q}^2)$. The dependence is rather weak, however, so that it is preferable to chose $c_M$ so that $\langle\lambda_{\max}(\hat{Q}^2)\rangle \approx 0.7$, probably depending on the polynomial used.

## 5 Combined Update

A further possible improvement, which was proposed in [4], is to update the gauge field and the bosonic fields simultaneously, giving a larger freedom to the gauge variable. When implemented in a local way, this however does not solve the $\tau_{\rm int} \propto n$ problem, since the gauge field is still coupled to $O(n)$ degrees of freedom which are kept fixed during the update. Still one can hope to decrease the proportionality constant considerably.

Our specific implementation of this idea is to update a gauge link while the bosonic fields at the end of the link are integrated out analytically. To be able to proceed to the next link we have then to revive these degrees of freedom by an exact update. The idea behind this is that the effective action for the gauge link after integrating the bosonic fields is still of the same form as the original action and the already existing updating methods for the gauge field apply.

To keep the formulas short, we do not write the index $k$ for the bosonic fields. Furthermore, we write

$$e^{-S(\underline{\phi},\ldots)} = \int \mathrm{d}\phi e^{-S(\phi,\ldots)}. \tag{5.1}$$

We propose the following transition probabilities for a three-step-update for the fields $U(x,\mu)$, $\phi(x)$ and $\phi(x+\hat{\mu})$:

$$P_1(U,\phi \to U',\phi') = N^{-1}\delta_{\phi(x)\phi(x)'}\delta_{\phi(x+\hat{\mu})\phi(x+\hat{\mu})'}e^{-S(U(x,\mu)',\underline{\phi(x)},\underline{\phi(x+\hat{\mu})})}, \tag{5.2}$$

$$P_2(U,\phi \to U',\phi') = N^{-1}\delta_{UU'}\delta_{\phi(x+\hat{\mu})\phi(x+\hat{\mu})'}e^{-S(U(x,\mu)',\phi(x)',\underline{\phi(x+\hat{\mu})})}, \tag{5.3}$$

$$P_3(U,\phi \to U',\phi') = N^{-1}\delta_{UU'}\delta_{\phi(x)\phi(x)'}e^{-S(U(x,\mu)',\phi(x)',\phi(x+\hat{\mu})')}, \tag{5.4}$$

where $N$ denotes the appropriate normalization factor (which can depend on $U(\mu,x)$ etc.). Alternatively, one can use an over-relaxation step for the gauge field, i.e. we have an operator $M$ which obeys $M^2 = 1$ and

$$e^{-S(M(U(x,\mu)),\underline{\phi(x)},\underline{\phi(x+\hat{\mu})})} = e^{-S(U(x,\mu),\underline{\phi(x)},\underline{\phi(x+\hat{\mu})})} \tag{5.5}$$

and $|DM/DU| = 1$ (for simplicity), then

$$P_1^o(U,\phi \to U',\phi') = \delta_{\phi(x)\phi(x)'}\delta_{\phi(x+\hat{\mu})\phi(x+\hat{\mu})'}\delta(U(x,\mu) - M(U(x,\mu)')) \tag{5.6}$$

also works. It is easy to show that both updates fulfill detailed balance. The combination of the three steps for the heatbath case is just

$$P_{123}(U,\phi \to U',\phi') = N^{-1} e^{S(U(x,\mu)',\phi(x)',\phi(x+\hat{\mu})')}, \tag{5.7}$$

namely a usual heatbath for all bosonic and gauge fields. For the over-relaxed algorithm, the effective updating probability is

$$P^{o}_{123}(U,\phi \to U',\phi') = N^{-1}\delta(U(x,\mu) - M(U(x,\mu)'))e^{-S(U(x,\mu)',\phi(x)',\phi(x+\hat{\mu})')}. \tag{5.8}$$

One can easily generalize the update to incorporate more fields, e.g. we can keep for example the site $\phi(x)$ integrated out while updating all four directions of $U(x,\mu)$ and $\phi(x+\hat{\mu})$. One should note that this update contains next-to-next-to-nearest neighbour interactions, which makes parallelization difficult.

Using some technical tricks described in appendix B, it is possible to make one sweep through the lattice only roughly 1.3 times slower than one standard iteration consisting of one bosonic heat-bath and two bosonic over-relaxation and two gauge over-relaxation sweeps. Not all tricks were implemented yet, therefore the autocorrelation times in units of $Q\phi$-operations may change by some percent.

We did some test runs on $4^3 \times 8$ lattices with $K = 0.165$ and $\beta = 1.75$ using the preconditioned action, the results of which we give in the following table:

| $n,\epsilon$ | $c_M$ | Updating | Plaq | $\tau_{\text{int}}(P)$ | $\tau_{\text{int}}(C_\pi(T/2))$ |
|---|---|---|---|---|---|
| 0.015 | 0.6 | a | 0.4708(15) | 20(5) | 18(8) |
| 16 | | b | 0.4720(14) | 17(4) | 10(1) |

The first line was obtained using one combined over-relaxed updating, in the second line, a bosonic over-relaxation step was added. On smaller lattices at smaller $K$ values a similar situation was found for the combined heatbath update. If we are very optimistic we could say that the efficiency of this update is comparable to that of the hybrid over-relaxation case. However, this update might have a different $z$ since the bosonic fields are always updated using a heatbath method. It would be preferrable to have a combined over-relaxation for both gauge and bosonic fields which travels the maximal allowed distance in phase space.

## 6 Application to other actions

All methods are easily applicable to the non-hermitian approximation proposed in [9] without substantial changes. Also the application to staggered fermions is straightforward. The improvements are also compatible with the global metropolis step proposed in [9,10]. One should note that in the metropolis step, when calculating the action difference, one has to use $P(\hat{Q})$ and not the transformed $\prod_k(\tilde{Q} - P_o r_k)(\tilde{Q} - P_o r_k^*)$.

More complicated is the case of the Sheikoleslami-Wohlert improved fermion action [16] which recently seems to attain more attention in unquenched simulations. There the fermion matrix has the form

$$M = \begin{pmatrix} D_{ee} & -KD_{eo} \\ -KD_{oe} & D_{oo} \end{pmatrix}, \tag{6.1}$$

where $D_{ee}$ and $D_{oo}$ are the clover terms $1 + Ki/2c_{\mathrm{SW}}\sigma_{\mu\nu}\mathcal{F}_{\mu\nu}$ which are diagonal in the position space indices. The preconditioned matrix takes the form

$$\hat{M} = D_{ee}(D_{oo} - K^2 D_{oe} D_{ee}^{-1} D_{eo}). \tag{6.2}$$

Here, $\gamma_5 \hat{M}$ is no longer hermitian, so we have to use a non-hermitian approximation. Note that $D_{ee}$ and $D_{oo}$ and its inverses are hermitian and commute with $\gamma_5$. Applying the same tricks as above, one can obtain the action

$$S_b = \sum \left|(M - \tilde{D}_{ee} r_k)\phi_k\right|^2 \tag{6.3}$$

with

$$\tilde{D}_{ee} = \begin{pmatrix} 0 & 0 \\ 0 & D_{ee}^{-1} \end{pmatrix}. \tag{6.4}$$

The bosonic updates are then easily calculated. The gauge update becomes rather cumbersome, since $D_{ee}$ contains gauge links and the there is no analytic expression for $D_{ee}^{-1}$. One has then to implement a numerical gauge over-relaxation step like the one described in [8]. A different possibility would be to handle $D_{ee}$ with an additional auxiliary field and use $\hat{M}' = D_{oo} - K^2 D_{oe} D_{ee}^{-1} D_{eo}$ as the preconditioned matrix. This matrix has the property that $\gamma_5 \hat{M}'$ is hermitian and allows us to use the hermitian approximation. We can then write

$$\det(D_{oo} - r_k - K^2 D_{oe} D_{ee}^{-1} D_{eo}) = \det \begin{pmatrix} 1 & -KD_{eo} \\ -KD_{oe} D_{ee}^{-1} & D_{oo} - r_k \end{pmatrix} \tag{6.5}$$

and, denoting the right hand side matrix with $\tilde{M}' - P_o r_k$, use the action

$$S_b' = \left|D_{ee}^{-1}\psi\right|^2 + \sum_k \left|(\tilde{M}' - P_o r_k)\phi_k\right|^2. \tag{6.6}$$

## 7 Conclusions

We have presented improvements of the local bosonic fermion algorithm proposed by M. Lüscher. A considerable speedup could be obtained by hybrid over-relaxation and preconditioning. Both algorithms were found to work well in a wide region of volumes. They are easily combined with other methods proposed in [6,9]. Also we were able to reduce the memory requirements by a factor of $\approx 3$. We found that the tuning of the parameters is not trivial and that the eigenvalues $\lambda_{\min}$ and $\lambda_{\max}$ need to be monitored. We found with $\delta = 3\%$ excellent agreement of the method with HMC and Kramers values [14] up to quite high statistics. The dependence on $\epsilon$ was, as expected, found to be small.

As for the case of the combined update, we found a comparable performance to the standard updates. There are still modifications which might work better and have to be investigated in the future. The update is compatible to the other improvements proposed so far. However, its implementation is tedious and is probably no longer applicable to Sheikoleslami-Wohlert improved fermion actions.

A performance comparison between optimized versions of the local bosonic method and the Kramers algorithm will be made in [15].

This algorithm still bears many possibilities of improvements, especially since its structure is relatively simple in comparison with HMC like methods. Its dynamics seems to be more easy to understand and allows for a more direct examination of its problems.

## 8 Acknowledgements

I wish to thank A. Galli, M. Lüscher, K. Jansen, S. Sint, R. Sommer and P. Weisz for many helpful discussions. I am also indebted to DESY Zeuthen for granting access to their computer resources.

## A Explicit formulae for the updates

Due to the similarity of the preconditioned action $\tilde{S}_b$ to the original one we can apply essentially the same techniques as for the original program, which were presented in [3] and, in great detail, in [2]. Here we will give the complete formulas for the preconditioned and combined case.

For convenience, we define $\chi_e(x)$ and $\chi_o(x)$ as 1 for $x$ on an even resp. odd site, 0 elsewhere. For the formulation of the bosonic updates in the preconditioned case, we write

$$\tilde{S}_b[U,\phi] = [\phi_k(x)]^\dagger A_k(x)\phi_k(x) + [B_k(x)]^\dagger \phi_k(x) + [\phi_k(x)]^\dagger B_k(x) + \text{constant}. \tag{A.1}$$

The coefficients $A_k$ and $B_k$ are easily evaluated:

$$A_k(x) = \tilde{c}_0^2(1+16K^2) + (r_k^* r_k - \tilde{c}_0(r_k^* + r_k)\gamma_5)\,\chi_o(x), \tag{A.2}$$

$$B_k(x) = [\tilde{Q}^2\phi_k](x) - \tilde{c}_0^2(1+16K^2)\phi_k(x) - \Big((\chi_e(x) r_k + \chi_o(x) r_k^*)([\tilde{Q}\phi_k](x) - \tilde{c}_0\gamma_5\phi_k(x)\Big)\,. \tag{A.3}$$

The heatbath updating can now be written as

$$\phi_k(x) \to A_k^{-1/2}(x)\chi - A_k^{-1}(x)B_k(x), \tag{A.4}$$

$\chi$ being a gaussian random $SU(2)$ spinor. The over-relaxation update is defined by

$$\phi_k(x) \to -\phi_k(x) - 2A_k^{-1}(x)B_k(x). \tag{A.5}$$

For the updates of the gauge field, we define two spinors $\xi$ and $\eta$ as

$$\xi = [\tilde{Q}\phi_k](x) + K\tilde{c}_0\gamma_5 U(x,\mu)(1-\gamma_\mu)\phi_k(y), \tag{A.6}$$

$$\eta = [\tilde{Q}\phi_k](y) + K\tilde{c}_0\gamma_5 U(x,\mu)^\dagger(1+\gamma_\mu)\phi_k(x). \tag{A.7}$$

They are independent of $U(x,\mu)$ and we can write

$$\begin{aligned}\mathrm{tr}\,\{U(x,\mu)F_k\} =& 2K\tilde{c}_0\mathrm{Re}\Big\{(\chi_o(x)r_k^* + \chi_e(x)r_k)[\phi_k(x)]^\dagger U(x,\mu)\gamma_5(1-\gamma_\mu)\phi_k(y) - \\ & \xi^\dagger U(x,\mu)\gamma_5(1-\gamma_\mu)\phi_k(y) - [\phi_x(x)]^\dagger U(x,\mu)\gamma_5(1-\gamma_\mu)\eta\Big\}.\end{aligned} \tag{A.8}$$

More compactly, the action is given by

$$\begin{aligned}\mathrm{tr}\,\{U(x,\mu)F_k\} =& v^\dagger U(x,\mu)w + w^\dagger U(x,\mu)v \\ =& \mathrm{tr}\Big\{U(x,\mu)v_k^\dagger w_k\Big\} + \text{h.c.}\end{aligned} \tag{A.9}$$

with

$$v = (1+\gamma_\mu)\left[\xi - \frac{\chi_o(x)r_k + \chi_e(x)r_k^*}{2}\phi_k(x)\right] - \frac{K\tilde{c}_0}{2}(1-\gamma_\mu)\gamma_5\phi_k(x), \tag{A.10}$$

$$w = (1-\gamma_\mu)\left[\eta - \frac{\chi_o(x)r_k^* + \chi_e(x)r_k}{2}\phi_k(y)\right] - \frac{K\tilde{c}_0}{2}(1+\gamma_\mu)\gamma_5\phi_k(y). \tag{A.11}$$

To use the standard updates we have to ensure that $F_k/||F_k||$ is an $SU(2)$ matrix. This can be done setting

$$\begin{aligned}(F_k)_{11} &= (H_k)_{11} + [(H_k)_{22}]^*, \\ (F_k)_{12} &= (H_k)_{12} - [(H_k)_{21}]^*\end{aligned} \tag{A.12}$$

with ($\alpha$ and $\beta$ are the color indices)

$$(H_k)_{\alpha\beta} = [(v_k)_\beta]^\dagger(w_k)_\alpha. \tag{A.13}$$

This works the same way when updating $SU(2)$ subgroups of $SU(3)$ matrices.

For the combined update, we can write the action in the following way:

$$\begin{aligned}\tilde{S}_b(U(x,\mu),\phi(x),\phi(x+\hat{\mu})) = & \frac{1}{2}\phi(x)^\dagger A_x\phi(x) + \frac{1}{2}\phi(x+\hat{\mu})^\dagger A_{x+\hat{\mu}}\phi(x+\hat{\mu}) + \\ & b_x^\dagger\phi(x) + c_x^\dagger U(x,\mu)^\dagger\phi(x) + \phi(x+\hat{\mu})^\dagger d_{x,\mu}^\dagger U(x,\mu)^\dagger\phi(x) + \\ & b'^\dagger_x\phi(x+\hat{\mu}) + c'^\dagger_x U(x,\mu)\phi(x+\hat{\mu}) + \text{h.c.} + \text{const.}\end{aligned} \tag{A.14}$$

where $A$, $b$, $b'$, $c$, $c'$ and $d$ do not depend on $U(x,\mu)$, $\phi(x)$ and $\phi(x+\hat{\mu})$. Then, the effective actions take the following form:

$$\begin{aligned}
\tilde{S}_b(U(x,\mu),\phi(x),\underline{\phi(x+\hat{\mu})}) =& \frac{1}{2}\phi(x)^\dagger \tilde{A}_{x,\mu}\phi(x) + \\
& \tilde{b}_x^\dagger\phi(x) + \tilde{c}_x^\dagger U(x,\mu)^\dagger\phi(x) - c'^\dagger_x A^{-1}_{x+\hat{\mu}} U(x,\mu) b'_x + \\
& \text{h.c.} + \text{const.} \qquad \text{(A.15)} \\
\tilde{S}_b(U(x,\mu),\underline{\phi(x)},\underline{\phi(x+\hat{\mu})}) =& -\tilde{c}_x^\dagger \tilde{A}^{-1}_{x,\mu} U(x,\mu)^\dagger \tilde{b}_x - c'^\dagger_x A^{-1}_{x+\hat{\mu}} U(x,\mu) b'_x + \text{h.c.} + \text{const.} \\
\equiv& \operatorname{tr}\left\{U(x,\mu)\tilde{F}_{\mu,x}\right\} + \text{const.} \qquad \text{(A.16)}
\end{aligned}$$

with

$$\begin{aligned}
\tilde{A}_{x,\mu} &= A_x - d_{x,\mu}A^{-1}_{x+\hat{\mu}} d^\dagger_{x,\mu}, \\
\tilde{b}_x &= b_x - d_{x,\mu}A^{-1}_{x+\hat{\mu}} c'x, \\
\tilde{c}_x &= c_x - d_{x,\mu}A^{-1}_{x+\hat{\mu}} b'x. \qquad \text{(A.17)}
\end{aligned}$$

$\tilde{F}_{\mu,x}$ can be calculated exactly as above. The updates then look as follows

$$\begin{aligned}
U(x,\mu) \to& U(x,\mu)' \quad (\text{over} - \text{relaxed or heatbath}), \qquad \text{(A.18)} \\
\phi(x) \to& \tilde{A}^{-1/2}_{x,\mu}\chi - \tilde{A}^{-1}_{x,\mu}\left(\tilde{b}_x + U(x,\mu)'\tilde{c}_x\right), \qquad \text{(A.19)} \\
\phi(x+\hat{\mu}) \to& A^{-1/2}_{x+\hat{\mu}}\chi - A^{-1}_{x+\hat{\mu}}\left(b'_x + U(x,\mu)'^\dagger c'_x + U(x,\mu)'^\dagger d^\dagger_{x,\mu}\phi(x)'\right). \qquad \text{(A.20)}
\end{aligned}$$

$\tilde{A}_{x,\mu}$ is no longer diagonal in the dirac indices, but its inversion and taking a square root can easily be done analytically using the properties of the dirac algebra. Explicit formulae for $b$, $c$ and $d$ are given for the preconditioned case by:

$$\begin{aligned}
d_\mu(x) &= -2\tilde{c}_0^2 K + (\chi_e(x) r_k + \chi_o(x) r_k^*)\tilde{c}_0 K\gamma_5(1-\gamma_\mu), \qquad \text{(A.21)} \\
c_\mu(x) &= -\tilde{c}_0 K(1+\gamma_\mu)(\gamma_5[Q\phi](x+\hat{\mu}) - \tilde{c}_0\phi(x+\hat{\mu}) + 2\tilde{c}_0 K U^\dagger(x,\mu)\phi(x)), \qquad \text{(A.22)} \\
c'_\mu(x) &= -\tilde{c}_0 K(1-\gamma_\mu)\left(\gamma_5[Q\phi](x) - \tilde{c}_0\phi(x) + 2\tilde{c}_0 K U_\mu(x)\phi(x+\hat{\mu})\right), \qquad \text{(A.23)} \\
b_\mu(x) &= B(x) - U(x,\mu)(c_\mu(x) + d_\mu(x)\phi(x+\hat{\mu})), \qquad \text{(A.24)} \\
b'_\mu(x) &= B(x+\hat{\mu}) - U^\dagger(x,\mu)(c'_\mu(x) + d^\dagger_\mu(x)\phi(x)). \qquad \text{(A.25)}
\end{aligned}$$

## B Implementation

All modifications were implemented on Quadrics Q1 and QH2 machines, which have a SIMD parallel architecture. For the updates, we apply the same tricks as already described in [2,3], namely we use the projector properties of $\gamma_5(1\pm\gamma_\mu)$ to minimize the number of $SU(2)$ times vector-multiplications. Furthermore, for the bosonic update of a field $\phi_k$ at a site $x$, the term $Q^2\phi_k(x)$ enters the calculation. We calculate and store $\psi = Q\phi_k$ before the sweep through the lattice and correct this vector after the update of each site $x$ and use $Q^2\phi_k(x) = Q\psi(x)$. This also saves a great amount of calculations. This trick can also be applied to the gauge and combined updates. Since here all $n$ fields enter for the update of one link, one has to store $Q\phi_k$ for all $k$ simultaneously. To avoid too large memory costs, one can store only slices of $Q\phi$. One can for example store $Q\phi$ for the timeslices $t-1, t$ and $t+1$ and then update the timeslice $t$ of the lattice. Proceeding to $t+1$, one has only to calculate $Q\phi$ at the slice $t+2$.

Another problem for the SIMD architecture is to update the lattice if points at local sublattices interact directly, i.e. when a point $x$ on the local sublattice on one node interacts directly with the point $x$ on the neighbouring node. This can happen when one divides the lattice into too small slices, i.e. when one wants to put an $8^4$ lattice on an $8^2 \times 4$ node cubic machine topology. This is even worse for the combined update, where also next-to-nearest neighbours interact, so that a sublattice has to have at least 3 sites in each direction. Since all processors have to work synchronously, interacting sites will be updates simultaneously, leading to wrong results. When one does not split one direction into slices, i.e. it is stored completely on each node, one can circumvent this problem. First we divide the processor array in even and odd nodes. Let us assume that we store the fields locally in an array `phi[x,y,z,t]` and that `T`, the temporal extent of the lattice, is even. On odd nodes, the same local coordinate can be stored in `phi[x,y,z,(t+T/2) mod T]`. Now, updating the same local index, one updates noninteracting physical sites. When coming to the boundary, one has, on even nodes, to add `T/2` to the time coordinate and on odd nodes subtract `T/2`. Modulo `T`, this is the *same* operation, therefore it is possible to do it synchronously, as required for SIMD machines.

## C Data analysis

Generally, autocorrelation times are expressed in units of $Q\phi$ operations to have comparable numbers for all updating schemes and lattice sizes. Multiplying the given numbers by the time needed by one $Q\phi$ operation one gets the autocorrelation time in units of seconds. However, we expect the $Q\phi$ units to be more portable to other machines since communication overhead effects and other effects like cache dependencies should be roughly equally present in a $Q\phi$ operation and the rest of the updates for most machine architectures.

For the data analysis several methods were applied. Masses were measured using the definition

$$m_{\pi,\rho} = \text{acosh}\left(\frac{C_{\pi,\rho}(T/2-1) + C_{\pi,\rho}(T/2+1)}{2C_{\pi,\rho}(T/2)}\right), \tag{C.1}$$

where $C_\pi$ and $C_\rho$ are the appropriate averaged meson correlation functions. No smearing was applied. Autocorrelation times for an observable $X$ were measured using the method proposed by Sokal [11], namely

$$\tau_{\text{int}} = \frac{1}{2} \sum_{t=-n+m}^{n-m} \frac{R(t)}{R(0)} \tag{C.2}$$

with

$$R(t) = \frac{1}{n-|t|} \sum_{i=1}^{n-|t|} (X_i - \bar{X})(X_{i+|t|} - \bar{X}) \tag{C.3}$$

and $m$ chosen so that $\tau_{\text{int}} \ll m \ll n$. We chose self-consistently the smallest value of $m$ for which $m/\tau_{\text{int}} \geq 4$. An estimate for the error of $\tau_{\text{int}}$ is given by

$$\sigma^2_{\tau_{\text{int}}} = \frac{2(2m+1)}{n} \tau^2_{\text{int}}. \tag{C.4}$$

For the remaining observables, the data was binned and analyzed using jackknife statistics. Then a plateau was searched in the estimate of the variance and this was taken as the estimate for the error. It was checked that this procedure gives compatible values to the formula $\sigma^2 = 2\tau_{\text{int}}\sigma^2_{\text{naive}}$. The bias found in the jackknife procedure for the masses was always at least two orders of magnitude smaller than the error. The thermalization was taken to be at least $10\tau_{\text{int}}$ iterations for systems with random start and $3\tau_{\text{int}}$ for pre-thermalized systems.

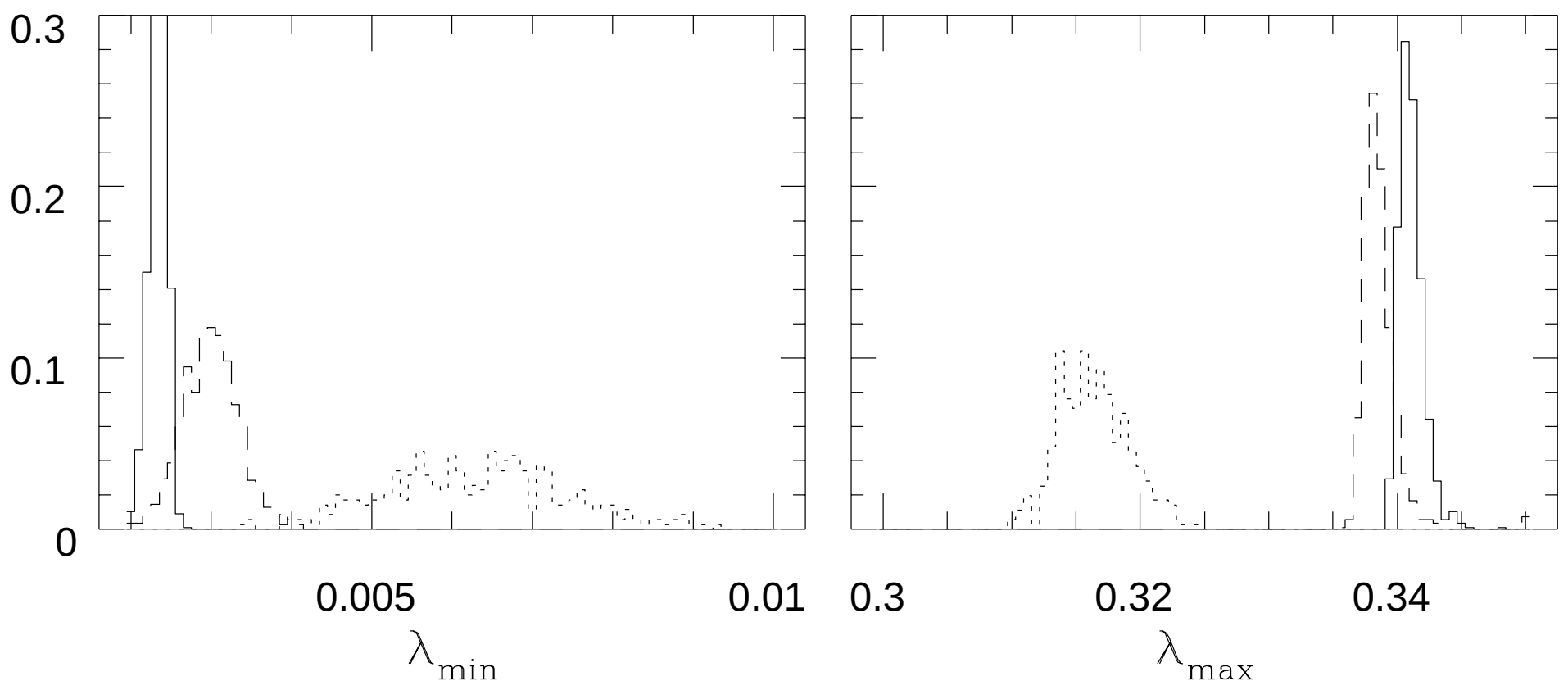

**Figure 1:** Distribution of the smallest (left) and largest (right) eigenvalues of $\hat{Q}^2$ for a $16^4$ (solid), $8^3 \times 12$ (dashed) and $4^3 \times 8$ (dotted) lattice.

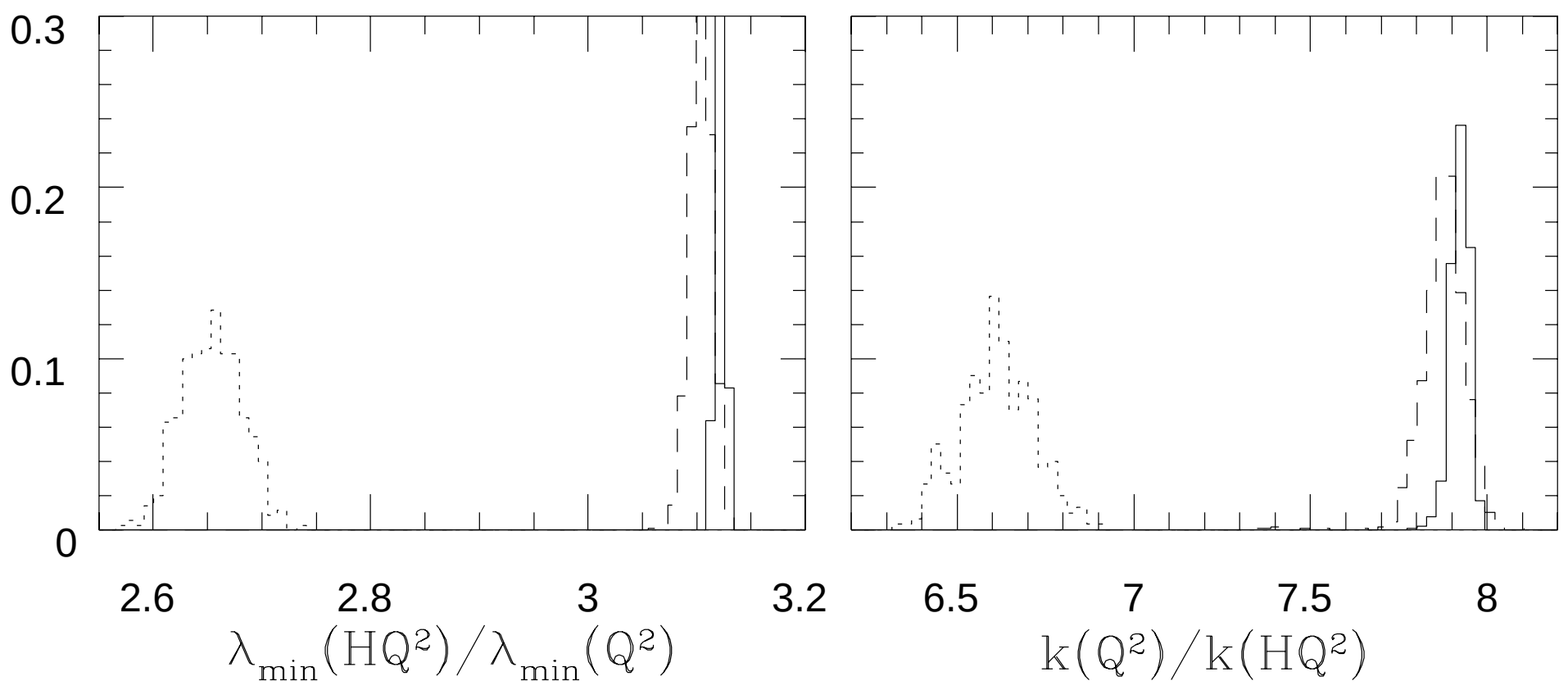

**Figure 2:** Distribution of $\lambda_{\min}(\hat{Q}^2)/\lambda_{\min}(Q^2)$ (left) and of the quotient of the condition numbers $k(Q^2)/k(\hat{Q}^2)$ (right) for a $16^4$ (solid), $8^3 \times 12$ (dashed) and $4^3 \times 8$ (dotted) lattice.

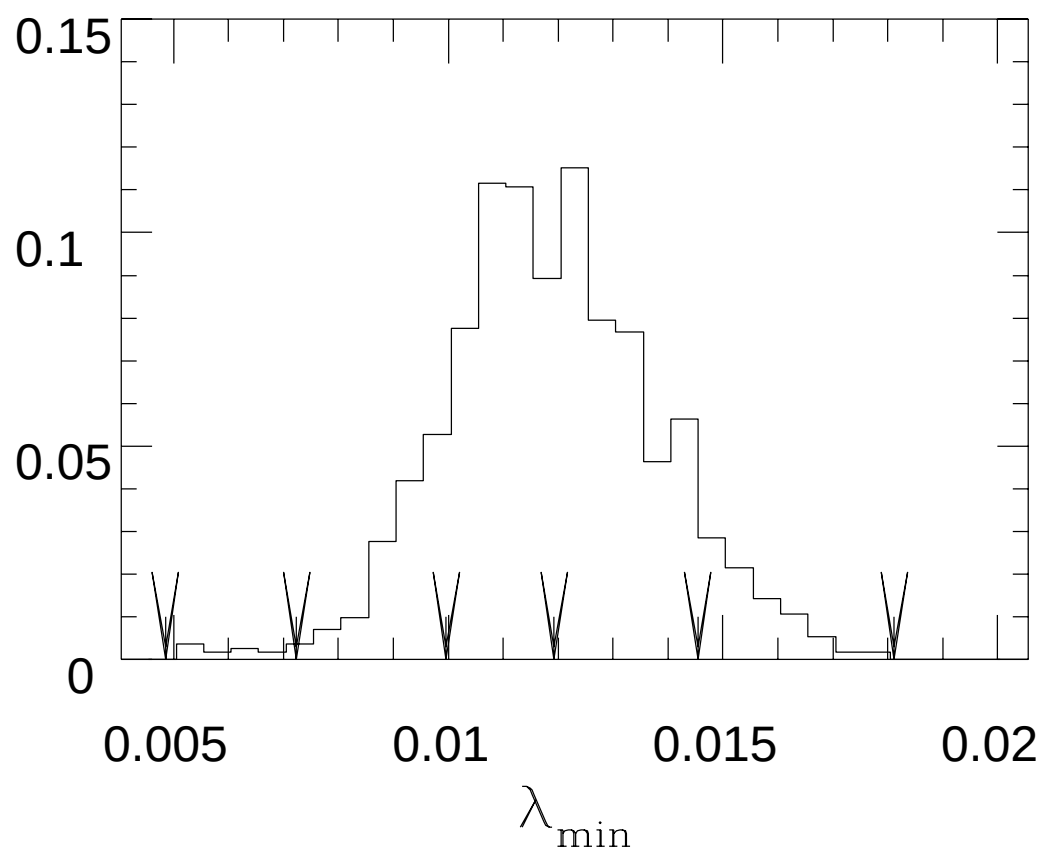

**Figure 3:** The locations of the $\epsilon$ values used in figure 6 in the distribution of $\lambda_{\rm min}(\hat{Q}^2)$

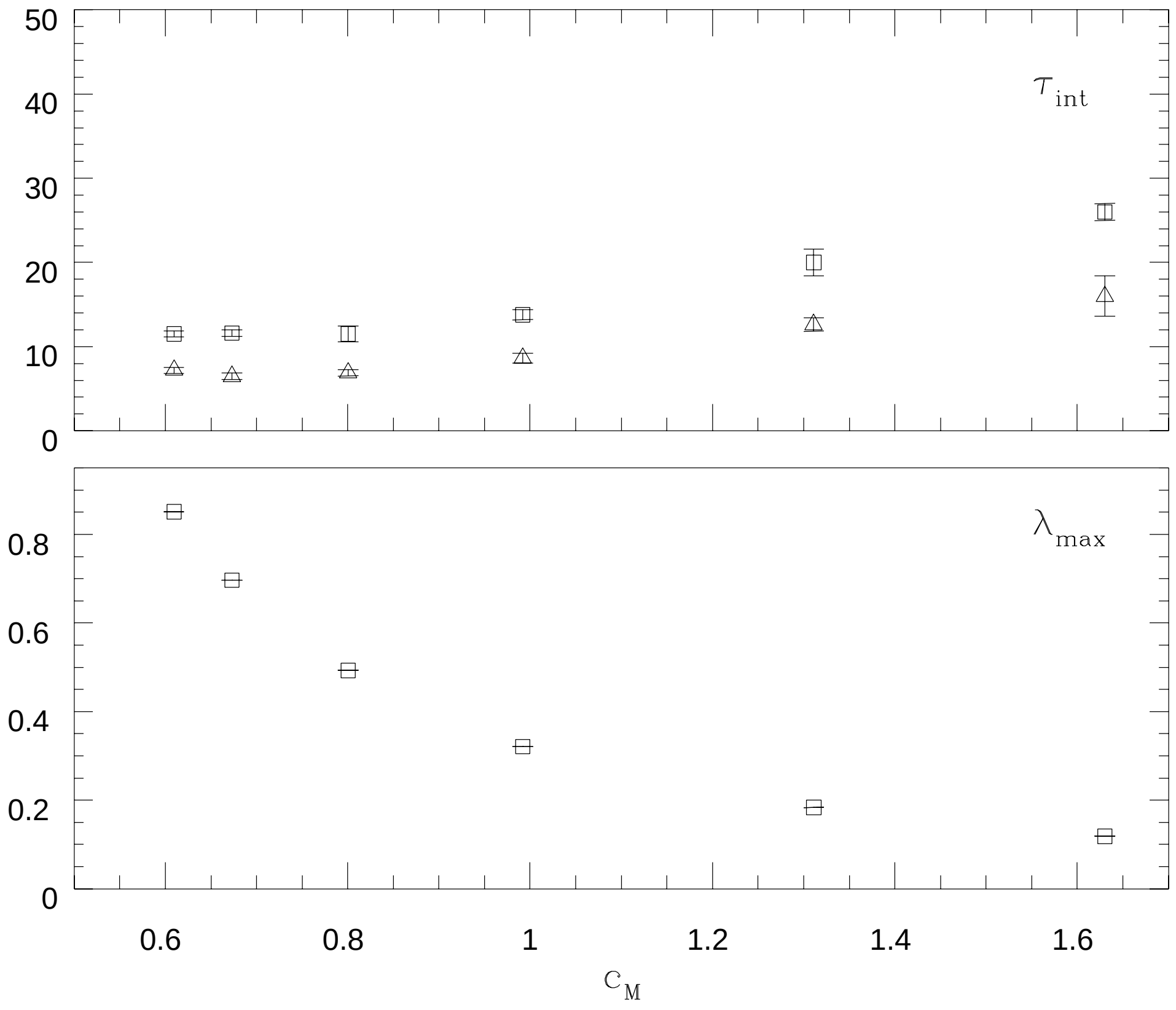

**Figure 4:** Dependence of the computational effort on $c_M$ when keeping $\delta$ and $\epsilon/\langle\lambda_{\rm min}(\hat{Q}^2)\rangle$ constant. The upper figure shows the computational effort in arbitrary units for the plaquette (boxes) and $C_\pi(T/2)$ (triangles), the lower figure $\langle\lambda_{\rm max}(\hat{Q}^2)\rangle$.

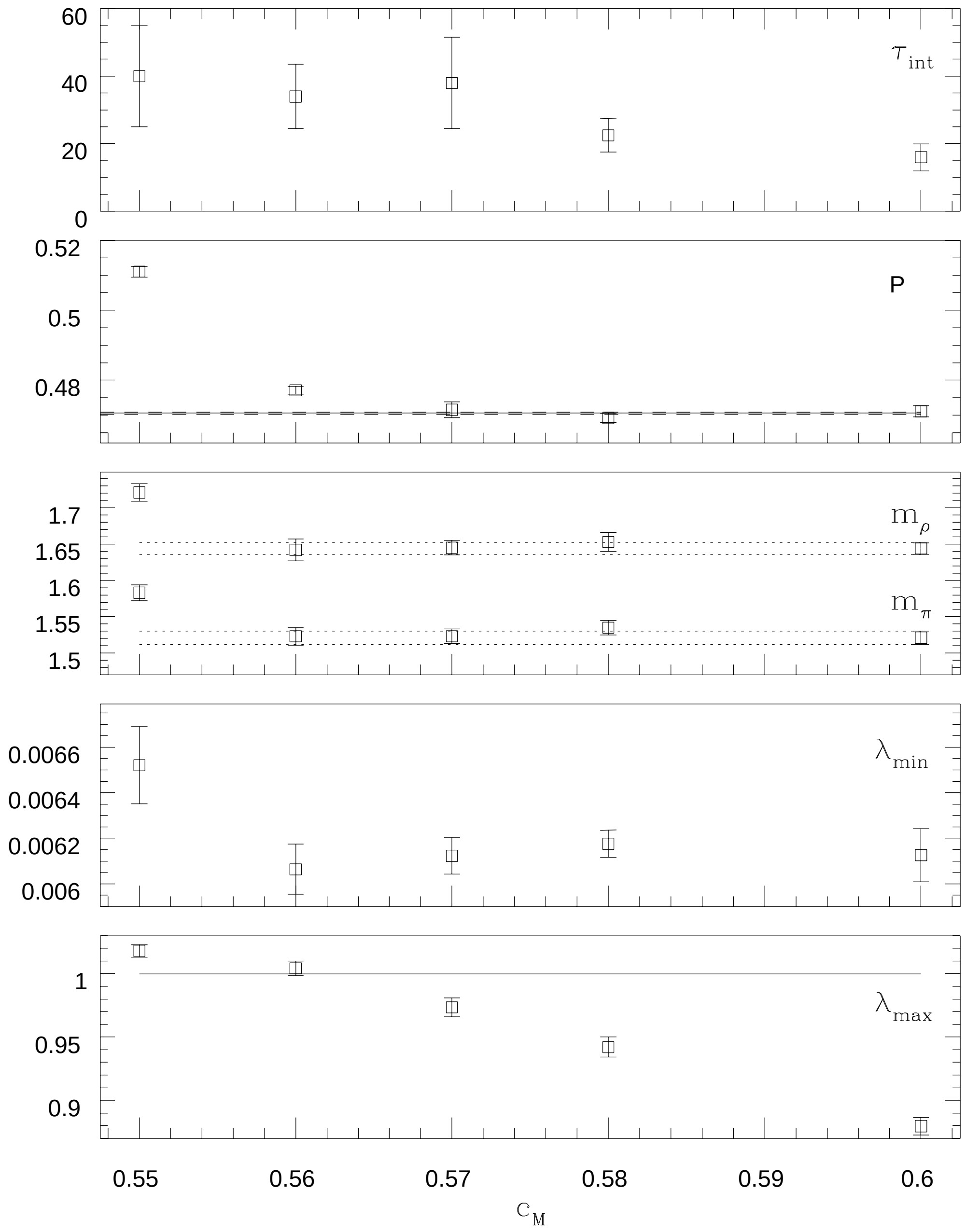

**Figure 5:** Dependence of several observables on $c_M$. From top to bottom there are $\tau_{\rm int}(\Box)$, the plaquette (the solid line gives the HMC value), the masses (here the $c_M = 0.6$ regions are shown as dotted lines), $\lambda_{\rm min}(\hat{Q}^2)$ and $\lambda_{\rm max}(\hat{Q}^2)$. Note that in the bottom figure the error bars give the width of the distribution of $\lambda_{\rm max}(\hat{Q}^2)$ rather than the errors.

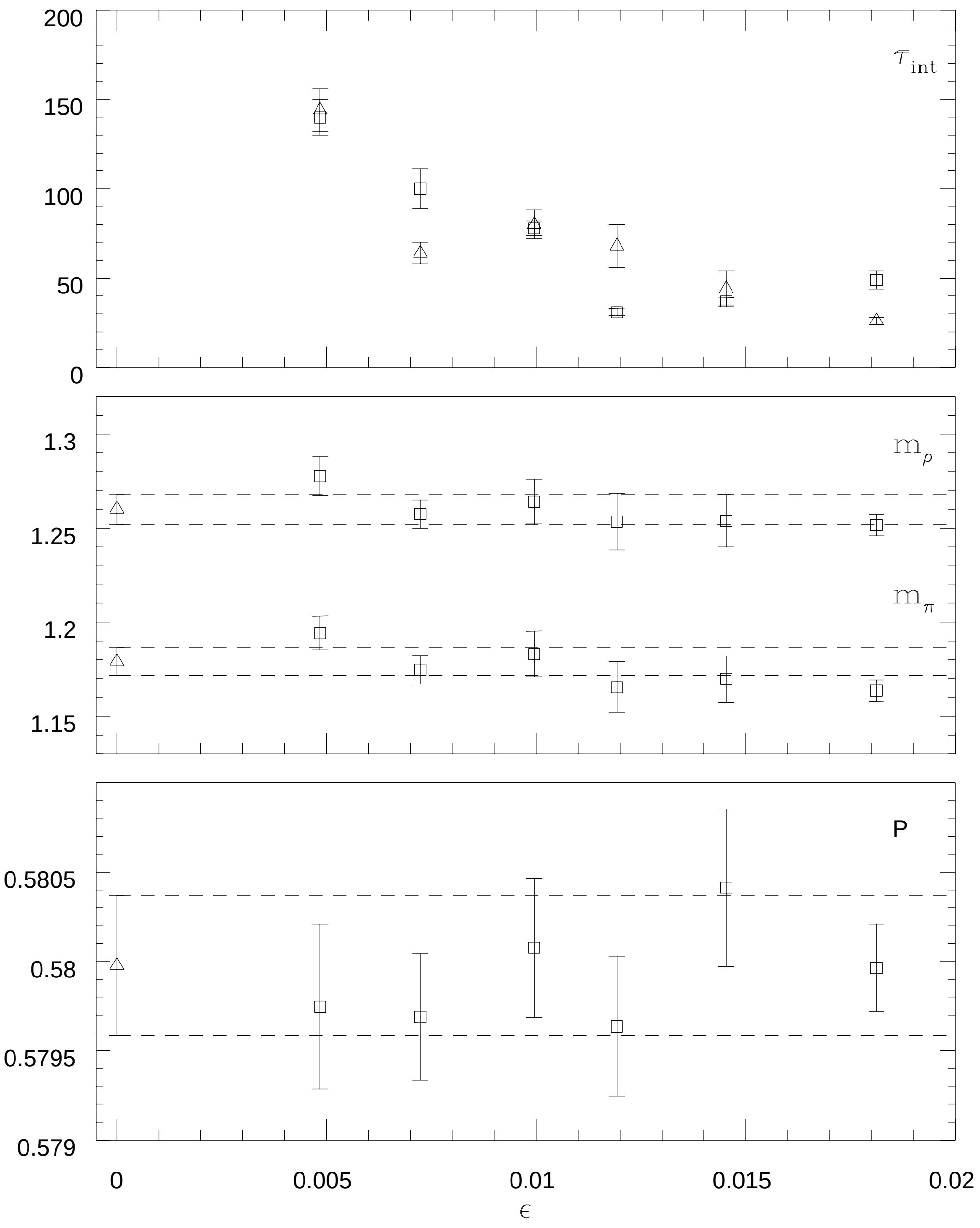

**Figure 6:** In the upper picture $\tau_{\rm int}$ for the plaquette (squares) and $C_\pi(0)$ (triangles) is shown. In the middle picture the $\pi$ and $\rho$ masses for the $6^3 \times 12$ lattice and several $\epsilon$ values are drawn. At $\epsilon = 0$ the HMC points are drawn with a triangle. The dashed lines indicate the HMC error bars. The lower picture shows the plaquette.